\documentclass [12pt,preprint]{aastex}

\begin{document}

\title{A multiple burst accretion Model to describe the metallicity 
distributions and stellar mass-metallicity relation for Local Dwarf Galaxies}

\author{F.D.A. Hartwick}

\affil{Department of Physics and Astronomy, \linebreak University of 
Victoria,
Victoria, BC, Canada, V8W 3P6}
\begin {abstract}

A one parameter model to describe the individual metallicity distributions 
and stellar mass-metallicity relation for dwarf galaxies is presented. 
This multiple-burst model is based on an accretion scenario, accomodates  
the observational constraint between $\overline{z}$ and $\sigma_{z}^{2}$ 
recently established by Leaman (2012), and predicts a slope consistent with 
the  stellar mass-metallicity relation of 
Kirby et al (2013) who showed that the local group dwarf spheroidal and dwarf 
irregular galaxies lie on the same relation. One interpretation of the model 
is that it describes star formation occuring either in gas 
rich mergers or at the intersection of colliding gas streams.

\end{abstract}


\section {Introduction}

Recently two important new results regarding the properties of dwarf galaxies 
have appeared. In one, Leaman (2012) showed that the same relation 
holds between the mean metallicity and the variance of the metallicity 
distributions for both the
dwarf spheroidal and dwarf irregular galaxies. Secondly, Kirby et al. (2013) 
have shown that the local group dwarf spheroidal and dwarf irregular galaxies 
lie on the same stellar mass-metallicity relation. This suggests that a common
evolutionary path 
exists for these two quite different morphological galaxy types. Clues to 
this evolutionary path must lie with the observed metallicity distributions. 
Unfortunately, the desired relationship is obscured because different 
evolutionary models can predict similar metallicity distributions e.g. Kirby 
et al. (2013) find that models with mass loss and those with accretion make 
quite similar predictions. Below we present an empirical model (the multiple-
burst model) which accomodates the Leaman (2012) constraint, predicts the 
slope of the stellar mass-metallicity relation consistent with the 
value determined by  Kirby et al.(2013) and is specified by only one 
parameter.

\section{The Model}

Kirby et al (2013) and earlier work have shown that the observed [Fe/H] 
distributions of the dwarf galaxies are nearly Gaussian-shaped functions with 
little or no metal poor tail. Both the simple closed box chemical evolution 
model (c.f. Searle and Sargent (1972), Pagel and Patchett (1975)) and the mass 
loss model of Hartwick (1976) predict a tail of metal poor stars which  
is more pronounced than in the above observations. Ways to limit the metal 
poor tail are to invoke either a pre-enrichment phase or an accretion model
\footnote{The accretion model used by Kirby et al. was designed by Lynden-Bell 
(1975) to model the metallicity distribution of the local G --
dwarfs.} 
(Kirby et al. 2013). Larson (1972) first introduced
the idea of infall (or accretion) in order to limit the number of stars in the
metal poor tail. Here, in addition to avoiding a metal poor tail, the desired 
model must maintain a body of enriched gas in order to account for the 
observations of the gas-rich dwarf irregular galaxies. We do this here by 
first generalizing the original Larson model. These results are then convolved
with a metallicity kernel to produce the multiple-burst model. The 
metallicity kernel is specified in order to accomodate the Leaman (2012) 
metallicity constraint. 

\subsection{Generalizing the Larson accretion model}

In the model of Larson (1972), star formation was assumed to 
occur at the same rate that gas was accreted. If $dM_{t}$ represents the 
incremental mass of gas accreted, and $dM_{s}$ and $dM_{g}$ represent the 
incremental mass of stars formed and gas accumulated respectively, then in the 
original model $dM_{s}/dM_{t}=1$ and $dM_{g}/dM_{t}=0$ i.e. $M_{g}=M_{g,0}=$ 
constant.

Here we parameterize the star formation rate and gas accumulation rate as 
$dM_{s}/dM_{t}=(1-q)$ and $dM_{g}/dM_{t}=q$ with $0<q<1$. When $q=0$
, the Larson (1972) result is recovered but now with $q \neq 0$ the star 
formation rate can be
lower than the accretion rate and enriched gas can be accumulated. 

In the interest of keeping the equations simple in what follows we will assume
that the symbols for metallicity $z$ and yield $p$ actually represent 
abundances with respect to the sun i.e. $z \equiv Z/Z_{\odot}$ and $p \equiv 
p/Z_{\odot}$.

The chemical evolution model used is the closed box model of Searle and 
Sargent (1972) and Pagel and Patchett (1975) where $M_{t}=M_{s}+M_{g}$ but 
here $dM_{t}\neq 0$ i.e. 
\begin{equation}
dz=1/M_{g}[(p-z)dM_{t}-pdM_{g}]
\end{equation}
where p represents the true nuclear yield. The equation is solved for $M_{g}$ 
by substituting $dM_{g}/q$ for $dM_{t}$ and integrating. Solutions for $M_{t}$
and $M_{s}$ then follow.
\begin{equation}
M_{g}=M_{g,0}/(1-z/((1-q)p))^q
\end{equation}
\begin{equation}
M_{t}=M_{g,0}/(q(1-z/((1-q)p))^q)-M_{g,0}(1-q)/q
\end{equation}
and
\begin{equation}
M_{s}=M_{t}-M_{g}=M_{g,0}(1-q)(1/(1-z/((1-q)p))^q-1)/q
\end{equation}
where we have assumed that the metallicity of the accreted gas is zero, and 
the initial amount of gas present is $M_{g,0}$. The metallicity distribution 
is easily obtained by differentiating the above expressions. Whereas in the 
Larson (1972) model most of the stars formed have $z=p$, here most of the 
stars and the accumulated gas have $z=(1-q)p$. A more compact, parameterized 
solution to equation (1) is given by Binney $\&$ Merrifield (1998).

The metallicity features predicted by the above model i.e. $dM_{s}/dz$ and $
dM_{g}/dz$ look like $\delta$ functions and as such do not resemble the 
observed distributions. Rather these features resemble the result of an 
instantaneous star burst. 
In order to proceed we shall assume that star formation takes place in many 
smaller such bursts possibly as a result of small fluctuations in the 
accretion rate and whose amplitudes are specified by a metallicity kernel 
$f(z)$. The final distributions are then obtained by a convolution of this 
metallicity kernel with the above generalized Larson model results e.g. the 
star formation burst occurs at $z=(1-q)p$ and the relative mass of 
stars formed in this burst $dM_{s}$ is $(1-q)f(z)dz$. Similarly for the other 
two components i.e. 
\begin{equation}
dM_{s}/dz=(1-q)f(z)=zf(z)/p
\end{equation}
\begin{equation}
dM_{g}/dz=qf(z)=(1-z/p)f(z)
\end{equation}
and
\begin{equation}
dM_{t}/dz=((1-q)+q)f(z)=f(z)
\end{equation}
The integration limits for all are $0 \rightarrow p$.

\subsection{The metallicity kernel}

The essence of this model is that chemical evolution evolution proceeds 
by repeated star bursts. The metallicity dependence is contained in the kernel.
 The perfect kernel needs to allow 
a description of the metallicity distributions while similtaneously providing 
a fit to the Leaman observations of $\overline{z}$ and $\sigma_{z}^{2}$ which 
is assumed to hold  over the full range of $z$ and the stellar 
mass-metallicity relation. Both of the latter conditions are integral 
constraints. Leaman describes the overall metallicity distributions with only 
two parameters: the mean and the variance. This means that in general the 
kernel within the convolution integrals is not unique.  The kernel chosen here
attempts to accommodate all three of the above conditions while maintaining  
mathematical simplicity. It contains only one parameter and allows for 
relatively simple evaluation of the convolution integrals. Any fine tuning to 
improve the fits will likely require more than one parameter and additional 
mathematical complexity. The following expression for the kernel is adopted.

\begin{equation}
f(z)=z^{2}e^{-z/a}
\end{equation}

Let the integrals of the moments of this function be represented as 
\begin{equation}
I_{n}=\int_{0}^{p}z^{n+1}e^{-z/a}dz=C_{n}(1-e^{-p/a}(1+g_{n}(p/a)))
\end{equation}
where $C_{n}$ is a function of a only and $g_{n}(p/a)$ is a polynomial in 
$p/a$.

the mean and variance of the stellar component is given by
\begin{equation}
\overline{z}=I_{3}/I_{2}
\end{equation}
and
\begin{equation}
\sigma_{z}^{2}=I_{4}/I_{2}-\overline{z}^{2}
\end{equation}

Note that when $a \ll p$, the integrals can be expressed in terms of the 
constants $C_{n}$ only. Under this condition the results of the model simplify
and the relation between $\overline{z}$ and $\sigma_{z}^{2}$ 
becomes 
$log \, \sigma_{z}^{2}=2 \, log\, \overline{z}-0.602$. The complete relation 
(for all
$p/a$) is shown superimposed on the observational data of Leaman (2012)  
in Fig 1. Both here and in what follows we have assumed that 
the yield is $p=1$ i.e. the solar value. Generally this model requires that 
the true yield be at least as high or higher than the the metallicity of the 
metal richest star under consideration.

In order to account for the results in Fig. 1 Leaman (2012) 
has proposed a scenario based on the chemical evolution model of Oey 
(2000). While one can expect that recurrent star bursts will cause 
fluctuations in the accretion rate, the solution above is empirically 
motivated. No 
attempt is made here to model the separate relation found by Leaman (2012) for 
the star clusters.

The stochastic nature of the model (recurrent star bursts) suggests that 
a Gaussian kernel might be appropriate. Models 
with a Gaussian kernel were constructed following the procedures given here 
and the results are described in the appendix.

\subsection{The model [Fe/H] distributions}

It is common practise to compare model [Fe/H] distributions with 
the observations in the $M_{s}-log(z)$ plane and to identify distributions by 
z at the maximum $M_{s}$ and to refer to this quantity as the effective yield 
$p_{eff}$. For the kernel we have chosen this quantity is $4a$ i.e. 
$a=p_{eff}/4$. From Eqn. (5) the appropriate distribution then becomes  
\begin{equation}
dM_{s}/d\,log\,z=\ln(10)\,z^{4}e^{-z/(p_{eff}/4)}/p
\end{equation}
This distribution is shown as a dotted line superimposed on the histograms of 
the observations of three dwarf spheroidal galaxies and two dwarf irregulars 
in Figs 2-6. The data for the dwarf spheroidal galaxies comes from Kirby et 
al. (2010) and for the dwarf irregular galaxies from Kirby et al. (2013).
The solid line in the figures is the dotted relation smoothed by 
a Gaussian kernel with the indicated $\sigma$ to show the effects of 
observational scatter. The fits are 'chi by eye'. All three distributions in 
each figure are similarly normalized. The fits are not perfect in that 
the model predicts a slight deficit of metal poor stars.

A related quantity of interest is the ratio of the mass in stars to the total 
mass accreted i.e. $M_{s}/M_{t}=I_{2}/(pI_{1})$.  When $a \ll p$ this ratio 
becomes $(3/4)p_{eff}/p$.

\subsection{The accumulated gas $M_{g}$}

Enriched gas accumulates because not all the gas accreted gets turned into 
stars. Assuming that this gas is well mixed, the mean metallicity $\overline{z}
_{M_{g}}$ is 
given by 
\begin{equation}
\overline{z}_{M_{g}}=(I_{2}-I_{3}/p)/(I_{1}-I_{2}/p)
\end{equation}
when $a \ll p$, this quantity approaches $\sim (3/4)p_{eff}(1-p_{eff}/4p)$ 
which is $\sim (3/4)\overline{z}_{M_{s}}$. 
It is reasonable to expect that some of this gas will be lost during the star 
bursts. However any that remains after the accretion phase ends should cool 
and as a result of any residual angular momentum form a disk. Stars may 
continue to form in this disk and if the object is sufficiently isolated will 
resemble a dwarf irregular galaxy. Dwarf spheroidal galaxies tend to be
found as companions to a central galaxy and currently do not have any gas 
remaining. They must have lost their accumulated gas by some process such as 
ram-pressure stripping as they fall into and orbit the halo of the central 
galaxy. This 
does not preclude additional star formation having occured prior to or even 
during the gas removal process however.

\subsection{The stellar mass-metallicity relation}

Kirby et al. (2013) have shown that the local dwarf spheroidal and dwarf 
irregular galaxies occupy the same stellar mass-metallicity relation. The 
relation between $\overline{z}$ and $M_{s}=I_{2}/p$ is easily constructed. 
When $a \ll p$ we note that $log\, \overline{z}=log\, a +log\, 4$ and that 
$log\, M_{s}=4\, log\, a -log\, p+log\, 6$ so that the slope of the relation 
is 
\begin{equation}
d\,log\,\overline{z}/d\,log\,M_{s}=0.25
\end{equation}

Kirby et al. (2013) find an observational slope of $0.30\pm 0.02$. The value 
predicted by this simple model is in reasonable agreement with the observed 
value.

The model predicts a total baryon mass-metallicity relation with a 
logarithmic slope ($d\,log\,\overline{z}/d\,log\,M_{t}$) of 0.33 when 
$a \ll p$.

\section{Discussion}
A multiple-burst accretion model of chemical evolution of dwarf galaxies is 
presented. The model
can acount for the individual observed metallicity distributions, the observed
mean z-variance relation, the observed stellar mass-metallicity relation and 
accounts  
for the distinction between dwarf spheroidal and dwarf irregular galaxies. 

The purpose of constructing simple models such as the one above is twofold: it 
provides a useful analytical description of the observations and it may 
provide some insight into the much more complicated processes associated with 
galaxy formation and evolution.  Given that the model is one of accretion 
suggests that star formation occurs incrementally with each incoming clump 
immediately turned into stars and enriched gas in a burst. This is in contrast 
with other models which start with a gas reservoir which is slowly converted 
into stars. The fact that not all of the gas is turned into stars suggests an 
inefficiency possibly the result of star formation occuring in gas cloud 
collisions rather than the less violent scenario of gravitational compression 
of gas within a stationary dark halo for example. The model may be describing 
what happens during early gas rich 
mergers or at the intersection of colliding gas streams. In fact this latter 
process could be responsible for the formation of the galaxies. More 
sophisticated modelling is required to validate the above interpretation. 

\section{Acknowledgements}

The author acknowledges financial support from an NSERC Discovery grant. He 
also thanks Adam Ritz for the use of his integral solver and Else 
Starkenburg and Ryan Leaman for discussions concerning the observations.

\section{Appendix}

As discussed above the stochastic nature of the model suggests that a Gaussian
kernel may be appropriate as is the case if the central limit theorem was to 
apply. We outline the solution with it here.

Let the kernel be defined as 
\begin{equation}
f(z)=e^{-(z-a)^{2}/2b}
\end{equation}
The convolution integrals become 
\begin{equation}
I_{n}=\int_{0}^{p}z^{n-1}e^{-(z-a)^{2}/2b}dz
\end{equation}
To illustrate the general form of the integrals the expression for $I_{2}$ is 
given by 
\begin{equation}
I_{2}=a\sqrt{\pi b/2}(erf(a/\sqrt{2b})-erf((a-p)/\sqrt{2b}))+b(e^{-a^{2}/2b}-
e^{-(a-p)^{2}/2b})
\end{equation}
where the error function $erf(x)=\frac{2}{\sqrt{\pi}}\int_
{0}^{x}e^{-t^{2}}dt$.

Letting $p_{eff}$ represent the maximum of the distribution in the $M_{s}-
log(z)$ plane requires the following relation between parameters a and b.
\begin{equation}
p_{eff}(p_{eff}-a)-2b=0
\end{equation}
Now let $a=\alpha p_{eff}$ so that $b=p_{eff}^{2}(1-\alpha)/2$.
Expressions for $\overline{z}$ and $\sigma_{z}^{2}$ can now be evaluated as 
above. The relation between $log \, \sigma_{z}^{2}$ and $log\, 
\overline{z}$ is well fit with $\alpha=0$ i.e. $a=0$ and $b=p_{eff}^{2}/2$ and 
the result is almost identical to that shown in Fig 1. When $p_{eff}/p
\ll 1$ this relation is $log \, \sigma_{z}^{2}=2 \, log\, \overline{z}-0.563$.

With $\alpha=0$ and $p_{eff}/p \ll 1$ the convolution integrals become very 
simple i.e $I_{1}=p_{eff}\sqrt{\pi}/2$, $I_{2}=p_{eff}^{2}/2$, $I_{3}=
p_{eff}^{3}\sqrt{\pi}/4$, and $I_{4}=p_{eff}^{4}/2$. Results in this regime 
follow easily i.e. $\overline{z}=p_{eff}\sqrt{\pi}/2$, $\sigma_{z}^{2}=p_{eff}
^{2}(1-\pi/4)$,  $\overline{z}_{M_{g}} \sim p_{eff}/\sqrt{\pi}$ and 
$d\,log\,\overline{z}/d\,log\,M_{s}=0.50$. This last result 
for the logarithmic slope of the stellar mass-metallicity relation is 
significantly higher than the  Kirby et al.(2013) result which is 
$0.30\pm 0.02$.

The [Fe/H] distribution is easily calculated as 
\begin{equation}
dM_{s}/d\,log\,z=\ln(10)\,z^{2}e^{-z^{2}/(p_{eff}^{2})}/p
\end{equation}

Fig. 7 shows this distribution calculated with the same parameters used in 
Fig.2. The two distributions are very similar although the Gaussian model is 
less symmetrical than the exponential model and may better account for the 
small metal poor tail. 

In summary, the physically motivated Gaussian kernel model satisfies very well
two of the three required 
conditions of the problem (the Leaman observations and the individual [Fe/H] 
distributions) but not the logarithmic slope of the stellar 
mass-metallicity relation.

\clearpage

\begin{figure}
\plotone{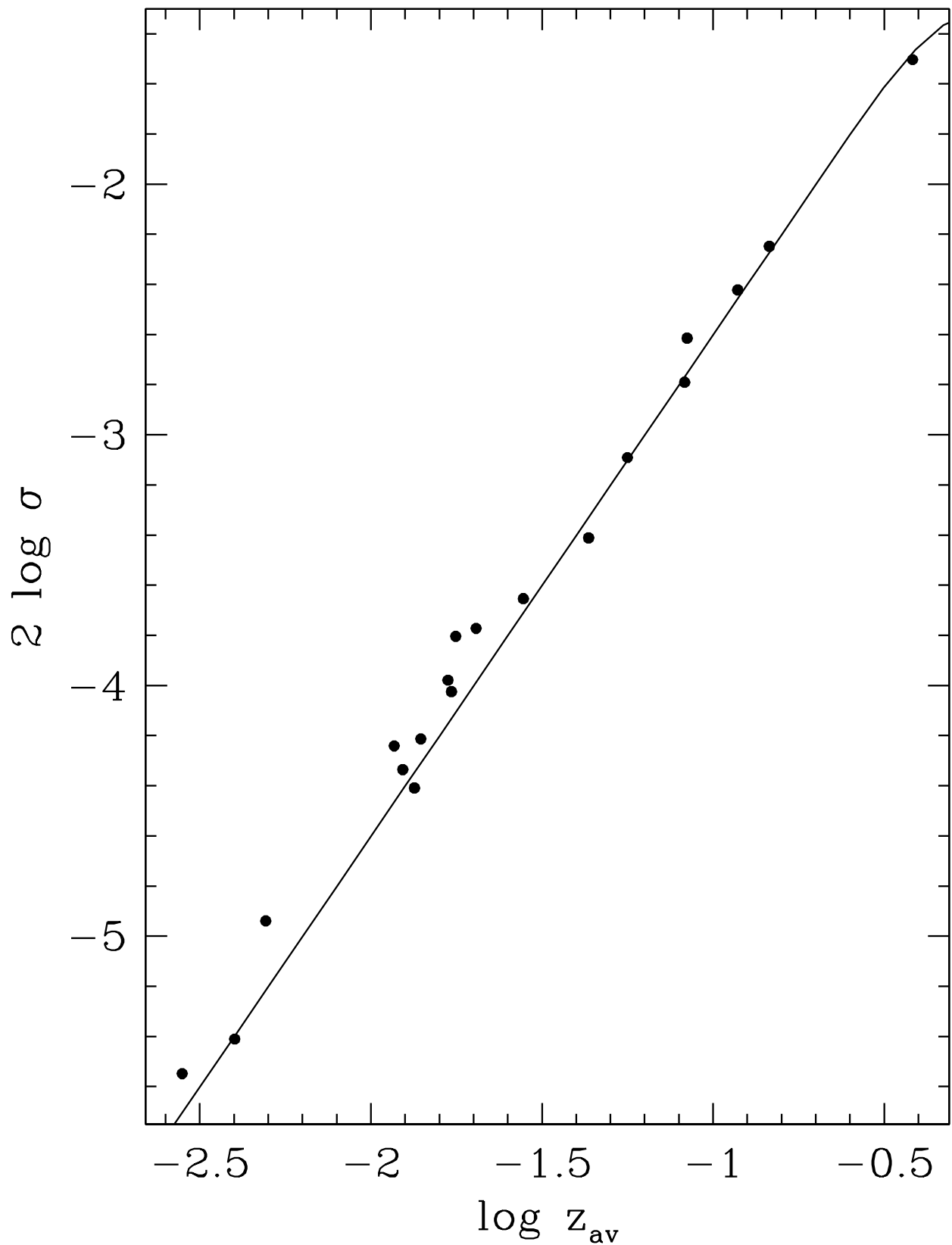}
\caption{Points -- the variance versus the mean metallicity of dwarf galaxies 
from Leaman (2012). The solid line is the relation calculated from the model.}
\end{figure}
\begin{figure}
\plotone{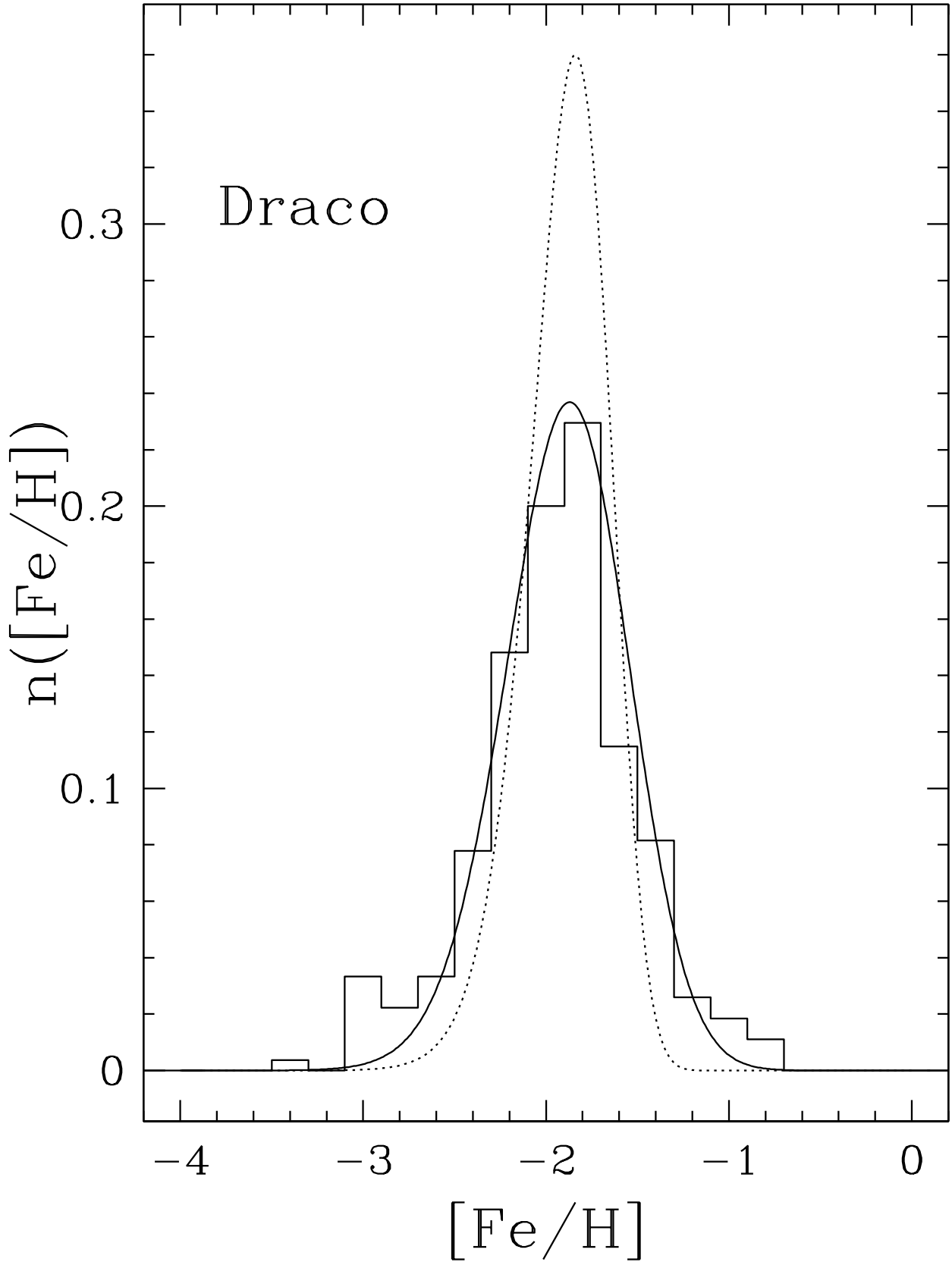}
\caption{[Fe/H] histogram for the Draco dwarf spheroidal galaxy. Dotted line 
-- the model distribution with $log \,p_{eff}=-1.84$. Solid line -- the model 
distribution smoothed by $\sigma=0.25$.}
\end{figure}
\begin{figure}
\plotone{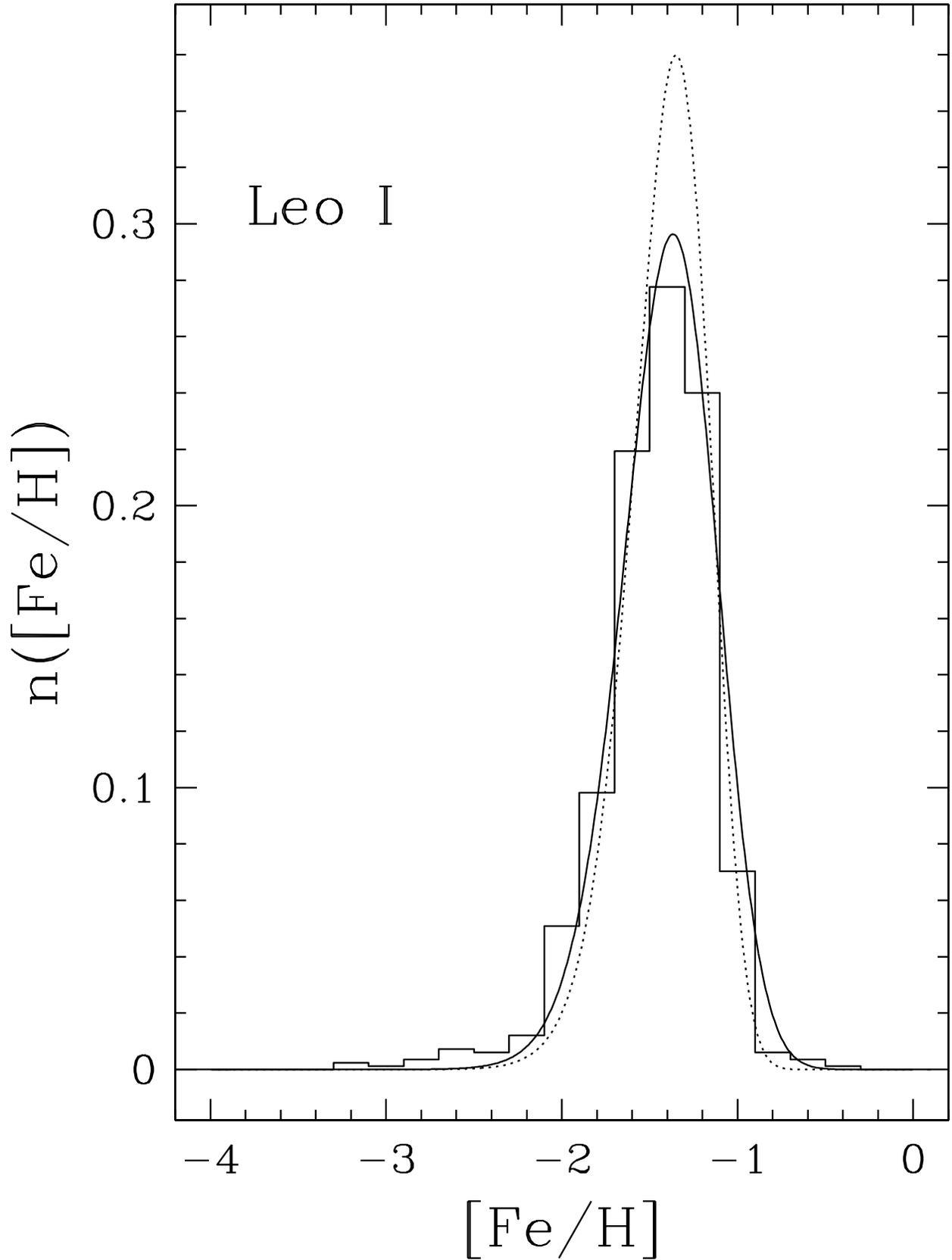}
\caption{[Fe/H] histogram for the Leo I dwarf spheroidal galaxy. Dotted line 
-- the model distribution with $log\,p_{eff}=-1.35$. Solid line -- the model 
distribution smoothed by $\sigma=0.15$.}
\end{figure}
\begin{figure}
\plotone{fig4.epsi}
\caption{[Fe/H] histogram for the Fornax dwarf spheroidal galaxy. Dotted line 
-- the model distribution with $log\,p_{eff}=-0.95$. Solid line -- the model 
distribution smoothed by $\sigma=0.15$.}
\end{figure}
\begin{figure}
\plotone{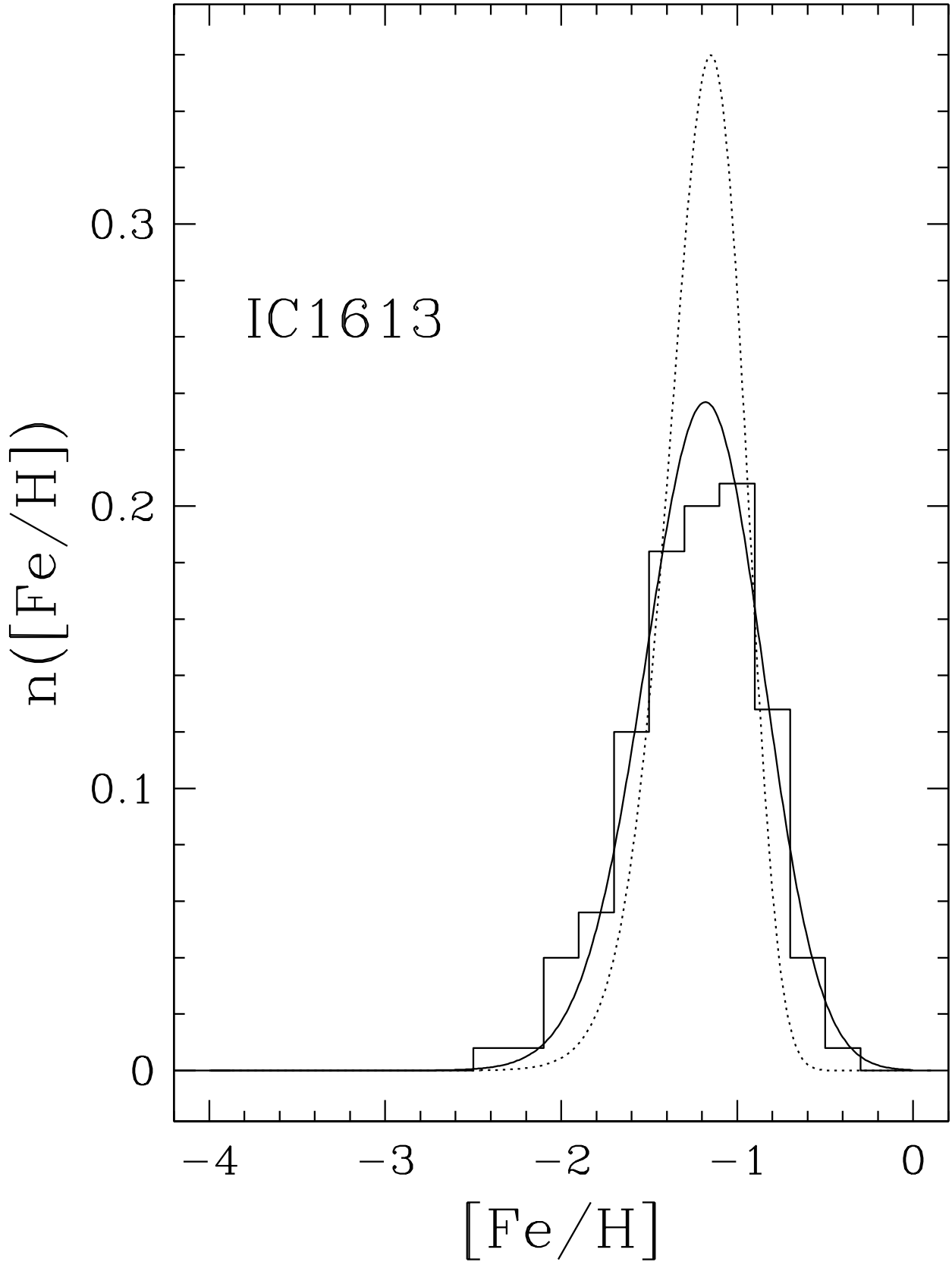}
\caption{[Fe/H] histogram for the dwarf irregular galaxy IC1613. Dotted line 
-- the model distribution with $log\,p_{eff}=-1.15$. Solid line -- the model 
distribution smoothed by $\sigma=0.25$.}
\end{figure}
\begin{figure}
\plotone{fig6.epsi}
\caption{[Fe/H] histogram for the dwarf irregular galaxy NGC6822. Dotted line 
-- the model distribution with $log\,p_{eff}=-0.95$. Solid line -- the model 
distribution smoothed by $\sigma=0.25$.}
\end{figure}
\begin{figure}
\plotone{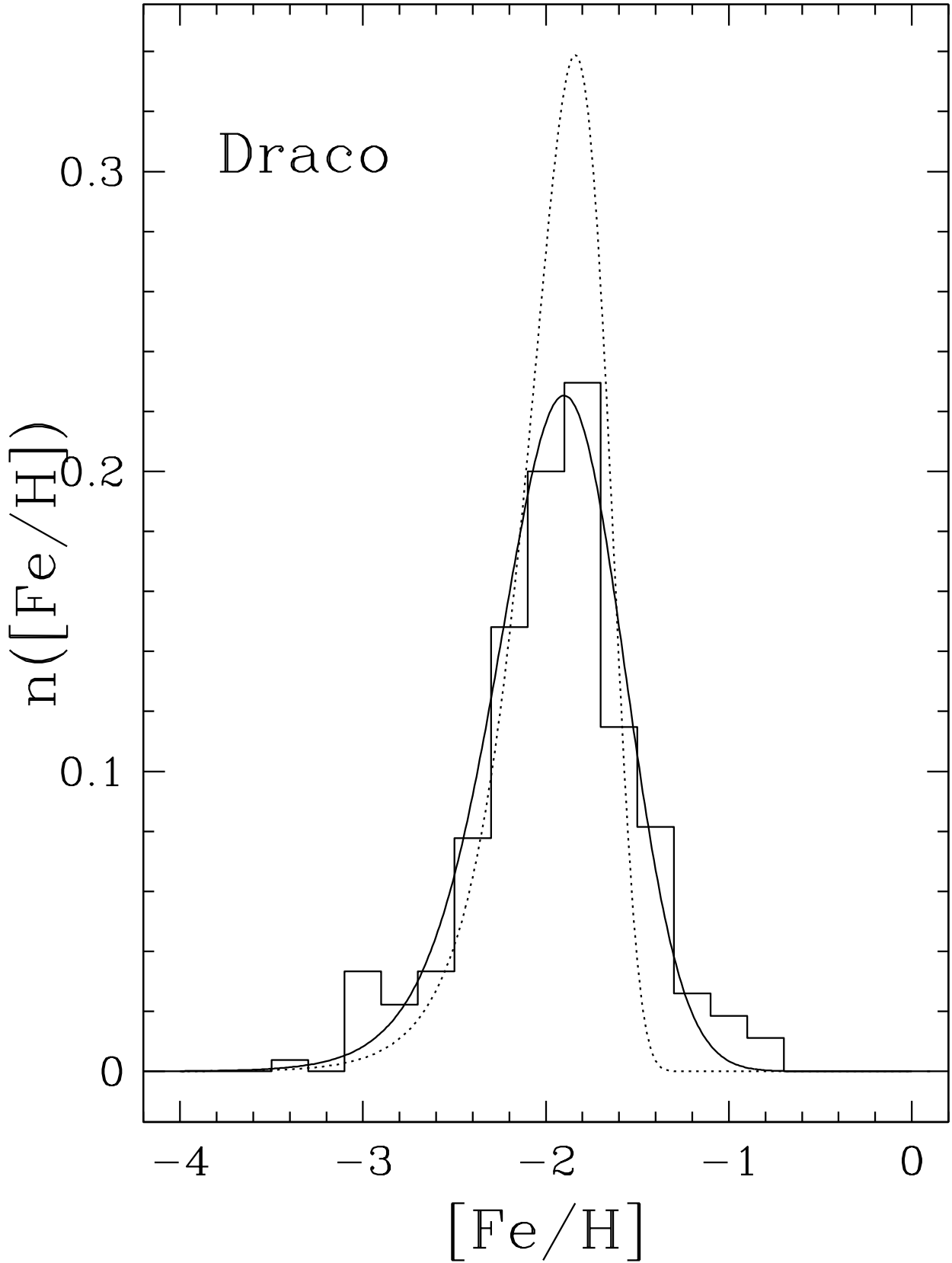}
\caption{[Fe/H] histogram for the Draco dwarf spheroidal galaxy. Dotted line 
-- the model distribution calculated with the Gaussian kernel with $log 
\,p_{eff}=-1.84$. Solid line -- the model distribution smoothed by 
$\sigma=0.25$. Compare with Fig.2}
\end{figure}

\begin{references}


\reference{}
Binney, J. \& Merrifield, M. 1998, Galactic Astronomy, (Princeton: Princeton 
University Press), 326
\reference{}
Hartwick, F. D. A. 1976, \apj, 209, 418
\reference{}
Kirby, Evan N.; Guhathakurta, Puragra; Simon, Joshua D.; Geha, Marla C.; 
Rockosi, Constance M.; Sneden, Christopher; Cohen, Judith G.; Sohn, Sangmo 
Tony; Majewski, Steven R.; Siegel, Michael, 2010, \apjs, 191, 352
\reference{}
Kirby, Evan N.; Cohen, Judith G.; Guhathakurta, Puragra; Cheng, Lucy; Bullock,
James S.; Gallazzi, Anna 2013, \apj, 179, 102
\reference{}
Larson, R. B. 1972, Nature Phys. Sci., 236, 7
\reference{}
Leaman, R. 2012, \aj, 144, 183
\reference{}
Lynden-Bell, D. 1975, VA, 19,299
\reference{}
Oey, M. S. 2000, \apj, 542, L25
\reference{}
Pagel, B. E. J. \& Patchett, B. E. 1975, \mnras, 172, 13
\reference{}
Searle, L. \& Sargent, W. L. W. 1972, \apj, 173, 25


\end{references}
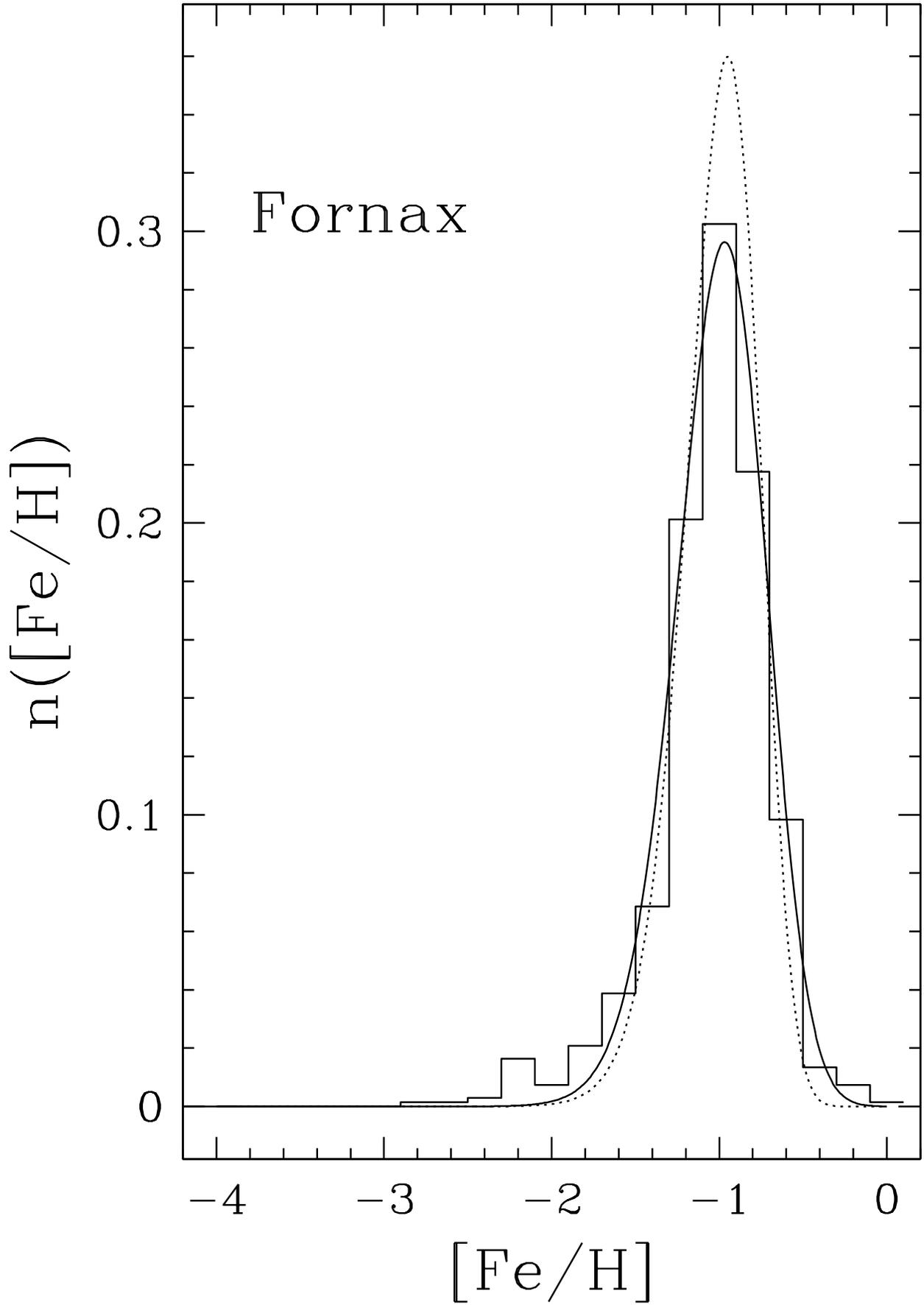
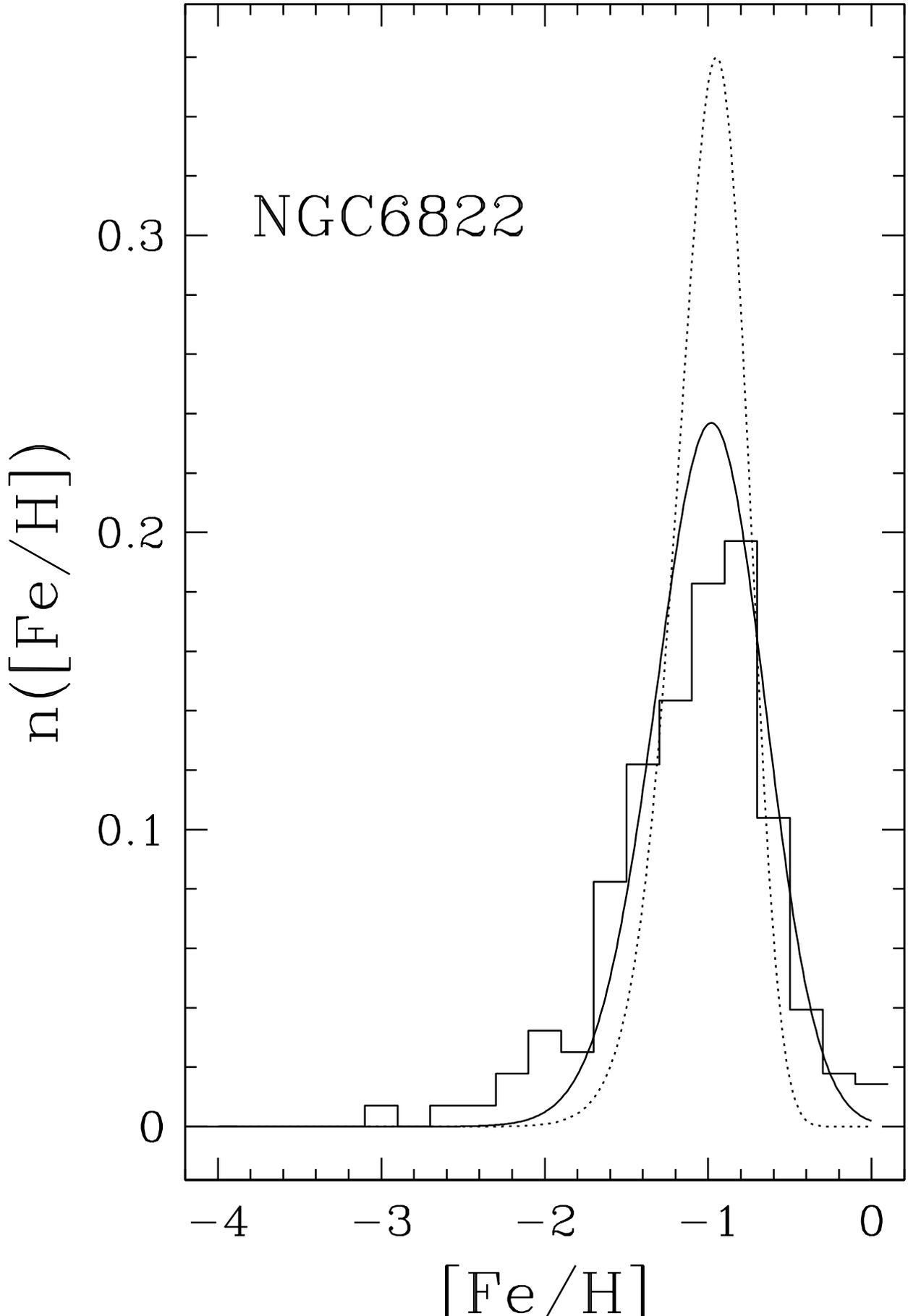
\end{document}